# Idealizations and Analogies: Explaining Critical Phenomena

**Quentin Rodriguez**

*Université Clermont Auvergne, PHIER, F-63000 Clermont-Ferrand, France.*
*E-mail address:* rodriguez.quentin@gmail.com

**Abstract**

The "universality" of critical phenomena is much discussed in philosophy of scientific explanation, idealizations and philosophy of physics. Lange and Reutlinger recently opposed Batterman concerning the role of some deliberate distortions in unifying a large class of phenomena, regardless of microscopic constitution. They argue for an essential explanatory role for "commonalities" rather than that of idealizations. Building on Batterman's insight, this article aims to show that assessing the differences between the universality of critical phenomena and two paradigmatic cases of "commonality strategy"—the ideal gas model and the harmonic oscillator model—is necessary to avoid the objections raised by Lange and Reutlinger. Taking these universal explanations as benchmarks for critical phenomena reveals the importance of the different roles played by analogies underlying the use of the models. A special combination of physical and formal analogies allows one to explain the epistemic autonomy of the universality of critical phenomena through an explicative loop.

## 1 Introduction

Critical phenomena (CP) and their property of "universality" have become a much discussed example in debates on the role of idealizations in scientific explanation and have been put forward when discussing reduction and emergence in physics. Batterman (2002a, 2002b, 2011) most notably linked these two aspects. His main insight, that Lange (2015) and Reutlinger (2017) challenge, is that some deliberate distortions (called "asymptotic reasonings" in Batterman, 2002a, or "minimal model explanations" in Batterman and Rice, 2014) are responsible for the capability of CP universality to unify the explanation of a large class of phenomena. This article aims to critically assess the argument behind this stance. Ultimately, I contend that the specificity of the explanation of CP universality is to be found in a particular combination of physical and formal analogies that produces an explicative loop, ensuring a kind of "epistemic autonomy."

In section 2, the main features of CP and their standard explanation, by means of renormalization group methods, are summarized. I discuss recent claims about the explanation of CP that oppose Batterman and favor a "commonality strategy" of explanation. In section 3, I systematically compare the explanatory strategy used for CP universality, the ideal gas model and the harmonic oscillator model, the latter two being construed as ideal types of model-based universal explanations in physics. The ideal gas model is an instance of micro-reductive universality, where physical analogies play an important role. Relatedly, the harmonic oscillator model is an instance of what I propose to call "Duhemian universality," for reasons explained below, where *formal* analogies are key. I then show how the specific combination of these two explanatory strategies can account for the epistemic autonomy of CP universality from underlying microscopic representations. Finally, in section 4, a discussion of this epistemic autonomy is provided in the context of the "commonality" debate.

## 2  What Is So Special About Critical Phenomena?

### 2.1  Critical phenomena and universality

In the last decades, CP in statistical and condensed matter physics have received much attention in the philosophy of scientific explanation and in the topic of emergence and reduction. This special type of phase transition, i.e., a sudden macroscopic reorganization of matter at thermodynamic equilibrium,[1] has drawn physicists' attention toward a specific feature they have named *universality*. Here, this term assumes a technical meaning: a property of a certain system is said to be *universal* if, in the vicinity of a phase transition, it behaves in the same way as other systems around their own phase transition, even if the microscopic constitution of these systems, the nature of their phase transition or the temperature of this phase transition are completely different.

For instance, the behavior of the fluids in the vicinity of their critical point is a CP. The liquid-gas phase transition is characterized by a discontinuous drop in density, since the liquid phase is denser than the vapor one; however, in the vicinity of the critical temperature and pressure, the difference in density between the two phases disappears and distinguishing the two states of the fluid becomes impossible. It has been experimentally shown that even if each fluid's critical points may be quite different depending on their composition,[2] the *curves* displayed by the main macroscopic quantities (such as density, specific heat capacity or compressibility) can be superimposed, up to a change of scale centered around the critical point (e.g., see Guggenheim, 1945, p. 257). These curves come to be power laws, each defined by a specific

---

[1] Mathematically speaking, a phase transition is usually defined by a non-analytical behavior of the free energy (the thermodynamic function macroscopically describing the system). CP are phase transitions where the first derivatives of the free energy are continuous.

[2] It can be different chemical elements, monatomic fluids like the noble gases, or complex molecular fluids like carbon dioxide. For instance, the critical temperature of xenon is 17 °C, and that of neon is *minus* 229 °C.

*critical exponent*. Taking the difference between the density of the liquid phase $\rho_l$ and the critical density $\rho_c$, we have:

$$\rho_l - \rho_c \sim \left|\frac{T - T_c}{T_c}\right|^\beta \qquad (1)$$

$T_c$ being the critical temperature and $\beta$ the critical exponent[3] (see Kadanoff et al., 1967, pp. 417ff. for an influential article, or the textbook Pathria and Beale, 2011, pp. 435ff.).

CP occur in many other systems, for instance in ferromagnetic materials (i.e., ferromagnets) or alloys. They all display quantities behaving like a power law in the vicinity of their critical phase transition, with a specific exponent shared by systems of different compositions. Moreover, the exponents of these different quantities are all related by a system of simple algebraic equations called *scaling laws*. For the critical point of fluids, $a$ being the critical exponent of the specific heat capacity and $\gamma$ that of compressibility, we have $a + 2\beta + \gamma = 2$. The system of scaling laws leaves only two independent critical exponents: giving two of them will set the others, no matter the CP.

Far more striking is that even systems of very different constitution may share the very same set of critical exponents, for completely different physical quantities. For instance, the difference in density between the two phases of a fluid near its critical point displays the same power law (Eq. 1), with the same critical exponent $\beta \approx 0.33$ as the spontaneous magnetization does in uniaxial ferromagnets near their Curie temperature. Here, the material undergoes its phase transition from paramagnetic (no spontaneous magnetization) to ferromagnetic (spontaneous magnetization produced by the magnet). Binary alloys, made of two chemical elements, also display CP when undergoing a transition between an ordered state (elements A and B are intertwined following a regular lattice) and a disordered state (probability of 0.5 to have element A or B on each site of the lattice). Noting $p$ as the statistical deviation from the disordered lattice, with $p = 1$ in a completely ordered lattice and $p = 0$ in a completely disordered lattice, it appears again to follow the same power law, with the same critical exponent $\beta \approx 0.33$, as our two previous examples (see Pathria and Beale, 2011, pp. 435ff.). Statistical physicists grouped such systems sharing the same set of critical exponents in *universality classes*, each universality class corresponding to the whole systems sharing one set of critical exponents. According to this classification, binary alloys near their ordered-disordered transition, fluids near their critical point and uniaxial ferromagnets near their Curie temperature belong to the same universality class. Other examples of CP (helium superfluidity, isotropic ferromagnets, etc.) exhibit *other* sets of critical exponents, and thus are associated with *other* universality classes than the latter one.

This is the empirical definition of a universality class. Now, how is this "universality" property actually *explained* by physicists? Without yet going into great detail, the standard explanation has been drawn from an empirical property of *scale invariance* since the 1970s, due to mathematical procedures using the *renormalization group* (RG) (for textbooks, see Goldenfeld, 1992; Binney et al., 1992; Lesne, 1998; for a brief overview, see Wilson, 1979; for an introduction for philosophers, Butterfield and Bouatta, 2015).

---

[3] $(T - T_c)/T_c$ is the reduced temperature used as a scale independent of the specific value of $T_c$ for each fluid. From a graphical standpoint, it corresponds to a rescaling of the curves to superimpose them. The symbol ~ means here "asymptotically equivalent, when $T$ is in the neighborhood of $T_c$, up to a constant multiplier." Equivalent definitions of $\beta$ can be found in the literature (e.g., taking the density of the gas phase instead of $\rho_c$).

This explanation usually starts from a highly idealized model, like the Ising model, that will be introduced more thoroughly in section 3. This mathematical model, made of "arrows" on a lattice with two possible states (up or down) having certain interactions between them, seems at first unsuited to quantitatively describe any real system. The Ising model was actually invented in the 1920s, before the formulation of quantum mechanics, to explain how the phenomenon of ferromagnetism could arise from "elementary micromagnets." This aim had been generally considered as a failure (see Niss, 2005 for a historical study). Indeed, the model is in contradiction with what has been later understood as a correct explanation of the quantum mechanical origin of ferromagnetism. It is now nevertheless used as a purely mathematical device to study phase transitions and CP, with or without any development on ferromagnetism itself.

The explanatory process takes advantage of scale invariance, i.e., the properties of the system are insensitive to certain dilations or contractions of scale. The fluid at its critical point, for instance, presents fluctuations of density of every possible length scale and appears as an indiscernible mix of vapor bubbles and liquid droplets inside one another. So, one performs an iterative "renormalization transformation" of the model, sweeping away all the complicated interaction details under a certain length, as if the model were described from a greater distance.[4] Taking the infinite limit of this operation, it can first be shown that every possible model sharing very few qualitative features with the one used as a starting point will flow to the same asymptotic limit under this renormalization transformation. This asymptotic limit is called a *fixed point* of the renormalization transformation, and its unstable directions determine these few features.[5] The abstract area in the space of the possible models, delimited by all the models flowing towards the same asymptotic limit, is called the *basin of attraction* of the fixed point.

Secondly, it can also be shown that the critical exponents of these possible models are only determined by mathematical properties of this fixed point (the eigenvalues of its unstable, or "relevant," directions). This result implies that these possible models have to share the same critical exponents and it allows their quantitative calculation. Finally, the scaling laws connecting the different critical exponents of each system, seen above, can also be proved.

Considering only systems with short-range interactions, there are essentially only two qualitative features that determine the critical fixed point towards which a certain model flows under the renormalization process and, subsequently, the critical exponents displayed by the systems behaving according to this model. The first feature is the number of space dimensions, $d$. For the three examples used here, we have $d = 3$. The second feature is the number $n$ of components of the *order parameter* which characterizes the phase transition ($n$ is sometimes called the symmetry number or the dimensionality of the order parameter). The order parameter is a macroscopic physical quantity that allows quantifying the transition and possesses a value of zero on the disordered side of the transition and a non-zero value on the other. For fluids it can be the difference in density seen earlier, for uniaxial ferromagnets the magnetization, and for binary alloys the statistical deviation from a disordered lattice $p$. These are all scalar quantities represented by one number ($n = 1$). Therefore, they all belong to the same universality class. The explanatory process outlined above shows that *any CP sharing the same (d, n) couple should then exhibit the same set of critical exponents* for a series of potentially completely different physical quantities. Thus, working with a toy model belonging to a universality class will enable us to *quantitatively predict* the behavior of complicated real systems near their phase transition, *even if* this model is completely unsuited to describe the systems considered. For the class discussed here, the Ising model is such an example. This eventually explains the enormous interest statistical physicists have shown for such very crude and unrealistic models. Goldenfeld thus contrasts such a "minimal model" viewpoint entailed by RG methods with a "'traditional' viewpoint," where the more accurate details the model contains, the better explanations it will provide (Goldenfeld, 1992, pp. 32–33).

---

[4] The iteration of the transformation needs to meet the properties of a semigroup. This explains why the method is called "renormalization group," even if, strictly speaking, it is not a *group*—the transformation is in general not invertible.

[5] The critical fixed point is thus a hyperbolic point, or "saddle point," stable when one crosses it along some direction, but unstable when one crosses it along another direction.

Here, let me clarify two different but sometimes confused explanations that can be considered in the context of CP. One can be interested in explaining why some critical phenomenon P displays some specific set of critical exponents E. The explanandum is then the value taken by the critical exponents of the phenomenon P, and the explanans will assign P to its universality class. The argument will typically look like this:

> P is a critical phenomenon characterized by a space dimension $d$ and an order parameter of $n$ components
> Any critical phenomenon of $d$ space dimensions and an order parameter of $n$ components will display the set of critical exponents E
> ---
> P displays the set of critical exponents E

The universal conditional of the explanans is the very definition of the universality class of P. Here, we have no explanation of the universality *itself*, since it is simply taken for granted in the explanans. The universality class is considered an empirical generalization, supported by experimental evidence. The novelty brought by RG methods we are interested in here has, however, been to explain the universality property itself, taken as an explanandum, i.e., the fact that an infinite class of real or imaginary systems *have to* share the same critical exponents, under very general hypotheses. This explanation cannot be put in such a strict deductive form (see for instance Palacios, 2019).

2.2 "A new way of looking at physics"

In brief, the main insight that this explanation of universality gives us may be summed up like so: to some extent, *even if* we do understand the critical behavior as a *pure result* of interactions between some elements (atoms, complex molecules, spins, etc.), we *do not need* to know these interactions and the nature of these elements to explain this global behavior.

Since the 1970s, this point has sounded genuinely new to theoretical physicists. Leo P. Kadanoff, one of the contributors to the development of RG methods, thus writes: "In one sense the renormalization group is rather different from anything that had come before in statistical physics, and by extension in other parts of physics as well." (Kadanoff, 2013, p. 179).

Robert Batterman suggested (Batterman, 2002a, 2002b, 2011; Batterman and Rice, 2014) that the crucial feature of the explanation of universality is rooted in the fact that one has principled reasons to deliberately eliminate details of the system under study, in order to identify some "universal" properties, precisely independent of these details. He dubbed this kind of operation *asymptotic reasoning* in Batterman (2002a), and RG methods entirely fit this idea. Characterized as such, RG methods are an instance of infinite idealizations, i.e., a deliberate distortion in which one parameter goes to infinity. In the case of CP, for instance, the account uses systems of infinite size and infinite number of particles (i.e., the thermodynamic limit). Then, we define a renormalization transformation on these systems, and take the number of iterations of this transformation to its infinite limit, as if the systems were replaced by how they appear "from an infinitely distant viewpoint." Batterman (2010) and Batterman and Rice (2014) have focused on the point that it is the *operation* of idealization *itself* that would be explanatory of universal behaviors, and not what is left *in spite* of these idealizations. This stand has been actively debated with proponents of "common features account" (or "commonality strategy") of explanation, like Marc Lange and Alexander Reutlinger (Lange, 2015; Reutlinger, 2017; reply in Batterman, 2019). To summarize, Lange and Reutlinger oppose Batterman and argue that the explanation of universality would be first rooted in the fact that an idealized model (e.g., the Ising model) shares features with the real systems belonging to its universality class. By approaching the debate from a new angle, I would like to argue here in defense of Batterman's insight, while giving credit to Lange and Reutlinger for some of their points.

This question may have been blurred by the conflation with a close but distinct debate— that of explanatory dispensability of infinite idealizations. This last debate has gained importance, since if these infinite idealizations are essential to the explanation, it can be considered as speaking in favor of the emergence of the explanandum, whereas if they are eliminable, it can be considered as speaking in favor of reduction (see Shech, 2018, for an overview). Indeed, a lot of attention physicists have given to CP and RG methods is about reductionism (e.g., see Laughlin and Pines, 2000; Brézin, 2004; Chibbaro, Rondoni and Vulpiani, 2014; for philosophical discussion Butterfield, 2011; Ardourel, 2018; Palacios, 2019). My goal here is more modest. I would like to take a step backwards, to focus on the previous question of the opposition between an explicative "commonality" versus an explicative "distortion" in the case of CP universality, irrespectively of the issue of an "in principle" dispensability of the infinite idealizations involved. I want to point out here what physicists consider as their best explanation of CP universality and examine what could be special in the way they explain this property in practice. Whether one could "in principle" get rid of these idealizations due to a "formally equivalent" reformulation of RG arguments does not dissolve the claims of strong novelty in the explanatory process currently used. Goldenfeld wrote for instance, in his textbook on CP, that "a new way of looking at physics emerged" (Goldenfeld, 1992, p. 16). It is not "a new physics," in the sense of a new field of phenomena, or "a new calculation tool" (that RG methods *also* are), but a new way of *doing* theoretical physics. Castiglione et al. (2008) made the same kind of claim: "The RG is not only a powerful calculational tool for determining asymptotic behaviors and their scaling exponents. It is also associated with a deep frameshift in our way of analyzing and modeling real phenomena." (p. 262)

2.3 Which kind of universality?

There seems to be a tension in the understanding of Batterman's claims about universality, which may be reflected in the debates surrounding them. The explanation of universality in CP by RG methods clearly constitutes his preferred example of "a certain type of reasoning that plays a role in understanding a wide range of physical phenomena" (Batterman, 2002a, p. 3). Even when discussing far simpler cases of explanation, like the Euler strut, the pendulum (Batterman, 2002a), or the Fisher's sex ratio model (Batterman and Rice, 2014), he keeps going back to CP, called the "*paradigmatic* example of universality" (Batterman, 2019, p. 27, emphasis in original). When he says that the explanation of CP universality "is an example of one of the best understood and most systematic explanations of universality" (Batterman, 2002a, p. 35), there is, however, a subtle shift in meaning about the word "universality" that might have been overlooked.

The notion of "universal" and "universality," in the common philosophers' meaning of being applicable to all individuals comprised in a class was already employed, in classical mechanics for instance, since at least the eighteenth century ("universal gravitation," "universality of the Newton laws," etc.). It became a specific and technical term only during the beginning of the 1970s, to qualify the precise phenomenon of the identity of critical exponents among large classes of systems, when explained by referring to scale invariance and renormalization.[6] No technical meaning was known for *universality* in the physicist community before, and there is no more precise definition of this notion among physicists. Its use has been extended since then, but only by analogy with CP (see Lesne, 1998), when a scale invariant property could have been identified, and the same RG mathematical apparatus could have been applied to obtain power laws (and their exponents) independent from the microscopic constitution of the phenomenon. Non-equilibrium phenomena called "dynamic critical phenomena," percolation, or protein folding, etc., have thus been assimilated to CP. Recently, a treatment of Hawking radiation of black holes as a "universal" phenomenon has also been suggested (see Gryb, Palacios and Thébault, 2019 for a philosophical comparison between CP universality and the proposed Hawking radiation universality). The extension of the notion is thus growing, but strictly speaking, physicists do not talk of universal properties in this technical meaning, *except* by using RG methods and by referring to CP universality as a template.

---

[6] See Fischer (1998): "the terms "universal" and "universality class" came into common usage only after 1974 when […] the concept of various types of renormalization-group fixed point had been well recognized […]. Kadanoff (1976) deserves credit not only for introducing and popularizing the terms but especially for emphasizing, refining, and extending the concepts." (p. 661). Also Domb (1985): "We have already mentioned empirical information accumulating about regularities in the pattern of critical exponents. These were collected together in the *hypothesis of universality* put forward by Kadanoff in 1970 at a summer school on critical phenomena (Kadanoff 1971), an analogous *smoothness postulate* was advanced independently by Griffiths (1970)." (p. 64, emphasis in the original text).

"Universality" can then be stretched between a restricted, technical meaning and a general, non-technical one. Wilson (1979) emphasized for instance this difference between CP universality and "trivial" universality, like gravitation and electromagnetic laws sharing the same exponent −2 (p. 174). The same can be said about the Van der Waals equation of fluids that incorrectly predicts the entire same equation of state for all fluids up to a rescaling of variables, instead of predicting only the identity of the exponents in the vicinity of the critical point shown by the actual data. Goldenfeld stressed that this would be "a form of universality […] quite different from the universality at the critical point" (Goldenfeld, 1992, p. 125).

However, Batterman does not seem to intend to study the explanation of CP universality for itself, but rather to use it as a paradigm example of a generic kind of explanation in physics ("asymptotic explanations" in Batterman, 2002a, or "minimal model explanations" in Batterman and Rice, 2014). Thus, in Batterman (2002a) the case of CP is used in order to exemplify "those features of asymptotic explanation that are likely *to generalize*" (p. 37), and, below, he explains that "It is possible to extract from this example [of CP universality] several *general features* that are present in *many explanations* of universality offered by physicists." (p. 42, emphasis mine on both quotations).

"Universality" is therefore intended to be understood in his work as an extended meaning, beyond CP universality. It clearly appears in this passage:

> While most discussions of universality and its explanation take place in the context of thermodynamics and statistical mechanics, we will see that universal behavior is really ubiquitous in science. Virtually any time one wants to explain some 'upper level' generalization, one is trying to explain a universal pattern of behavior. (p. 4)

This generalized universality would be nothing new in physics: "[…] I believe that such methods (broadly construed) are far more widespread in the history of science than is commonly realized. They play important roles in many less technical contexts [than RG methods]." (p. 7). Among these less technical contexts, we can even find one of the simplest systems considered in physics—the pendulum:

> 'Universality,' as I have noted, is the technical term for an everyday feature of the world—namely, that in certain circumstances distinct types of systems exhibit similar behaviors. (This can be as simple as the fact that pendulums of very different microstructural constitutions all have periods proportional to the square root of their length. […]) (p. 9, the example of the pendulum is detailed pp. 13–17)

In the more recent Batterman and Rice (2014), he tries to extend this "broadly construed" universality to biology, in analyzing the Fisher's 1:1 sex ratio model through the lens of CP universality and RG methods, even if for now neither an argument has been actually drawn from scale invariance nor a comparable mathematical apparatus has actually been applied to this example.

Batterman's "broadly construed" meaning of universality thus lies somewhere between the common, non-technical meaning (as when one says that gravity is a universal law) and the technical universality. In this article, I will focus instead on the latter. I am not trying to discuss the relevance of this heuristic extension of the notion of universality "broadly construed," and the suggestions Batterman draws from it. I only want to point out that one can disagree about what is considered crucial in the explanation of CP universality, if this case is implicitly considered as exemplifying different extensions of the notion of "universality." Such a clarification could prevent misunderstandings that might occur in the debate. If universality "broadly construed" includes the traditional explanation of the behavior of the simple pendulum, as Batterman suggests, then clearly it is a notion whose mesh is too wide to hold back the answer to the question raised above, that is: what would be new in the way physicists use RG methods to explain CP universality? One day, all these broadly construed "universal" phenomena, including the behavior of pendulums and the 1:1 population sex ratio, might be explained in the same way as CP, thanks to an appeal to a scale invariant property and RG methods, thus showing a deep connection between these explanations. But I do not think it is the case *today*, and this leaves our question of the claimed strong specificity of the explanation of CP universality unanswered.

2.4 Is universality just explained by a "commonality strategy?"

This clarification of what is at stake when we are talking about the explanation of universality can help in the ongoing debate between Batterman and proponents of the "common features account" of explanation, as he called them (Batterman and Rice, 2014), which Alexander Reutlinger prefers to call the "commonality strategy" (Reutlinger, 2017). Let us recall that in order to perform RG methods leading to the demonstration of the universality of critical exponents, we have to start with a strongly idealized model, like the Ising model. The RG methods show us that almost all possible details of the model are irrelevant to the determination of its critical exponents, *except* a few very generic features—essentially, the dimension of space $d$ and the number of components of the order parameter $n$—and allows us to calculate these critical exponents. If we then state that a given real system displays the same generic features as the idealized model, we can put it in the same universality class as the model, and finally predict its critical exponents without further investigation.

Still, Batterman strongly denies that these common generic features (the parameters ($d$, $n$) of the universality class), shared by the idealized model and some real systems, essentially explain the universality of critical exponents. Elaborating on other examples of explanatory strategies supposedly similar to that of CP, Batterman and Rice say for instance, "We contend that what accounts for the explanatory power of many of these caricature models is not that they accurately mirror, map, track, or otherwise represent real systems." (Batterman and Rice, 2014, p. 373). In another article, Batterman argues that "Explanations do not necessarily have to be representative. And, in many (most) instances of explanations in applied mathematics, they are not." (Batterman, 2010, p. 23).

This conviction probably originates in the passage from Goldenfeld (1992, pp. 32–33), quoted in section 2.1, according to which RG methods have supported a counterintuitive "viewpoint" on strongly idealized models, that Goldenfeld calls "minimal models." To put it simply, *the more* commonalities between a model and some phenomena there are, *the less* the model can be used to produce explanations; *the more* the model is stripped of properties, *the more* one may use it to explain. Of course, the simpler a model is, the more tractable it is. This is not the point considered here, which is far more specific. According to statistical physicists, RG methods provide a *theoretical justification* to say that some idealized model does not need to (more or less) *represent*—in the sense of sharing more or less common features with a given real system—to allow an explanation and an accurate calculation of critical exponents. A richer model would be *less* relevant for the explanation. This "deep frameshift" (Castiglione et al., 2008, quoted section 2.1) in considering the role of models in explanation is at the root of the idea of a "new way of looking at physics" I want to focus on.

By contrast, Reutlinger has argued that the common features belonging to both the idealized model and the real systems *are* explanatory, and that "RG explanations are not special and quite intuitive in [that] crucial respect" (Reutlinger, 2017, p. 144). Marc Lange has expressed the same opposition to Batterman's claim, although discussing other examples than CP: "Since the model's explanatory utility arises from its having certain features in common with the target system without the disturbing factors present in the target system, the model's explanatory utility arises from its representing accurately enough the target system, contrary to [Batterman and Rice]'s view." (Lange, 2015, p. 298)

There is probably a need to clarify the assessed explanandum, as Batterman (2019, pp. 41–43) seems to reply to Reutlinger and Lange. As discussed in section 2.1, to explain (1) why certain real systems display some set of critical exponents, is not the same question as to explain (2) why the class of the infinity of real or imaginary systems that share the same exponents, i.e., the universality class, is defined only by the parameters ($d$, $n$). Saying that fluids display a certain set of critical exponents *because* the Ising model possesses the same parameters (3, 1) as the critical transition of the fluids and displays such a set of critical exponents, is answering the question (1) using as an explanans the answer to the question (2). Since that question (2) is more general than the question (1), this makes no difficulty. The difficult but interesting part is that RG methods start from *a particular idealized model* as for answering the question (1), and in the course of explaining (1), they answer *as well* the question (2), thus justifying the initial use of the given idealized model, in a stronger sense than this model simply sharing partial similarity with the real systems considered.

Thus, I take the core of Batterman's intuition here to be that there is *something* gained in the process of explaining CP universality that "jumps above" the possibility of one enriching the idealized model to make it closer and closer to a supposedly "faithful" representation. I share this idea, but still, Reutlinger has raised a very interesting point of criticism, asking for the difference between the explanation of CP universality and that of the ideal gas law, a "paradigm of the commonality strategy" (Reutlinger, 2017, p. 148). All gases, at large dilutions, behave according to the same law relating pressure $p$, volume $V$, the amount of substance $n$ and absolute temperature *T* of the gas: the famous "ideal gas law" *pV* = *nRT* (*R* being the ideal gas constant). Broadly speaking (and *not* in the technical meaning), this law can be considered as a case of "universality," usually explained thanks to statistical mechanics. We start from a simplified microscopic description of a specific gas (called the ideal gas model), with a specific chemical composition and molecular mass, and we end up with a single relation, valid for any gas, no matter the precise nature of its constituents (the process is detailed in section 3.1). We can thus show how one universal behavior results from the diversity of real gases, independent of their detailed differences. One indispensable element of this explanation, highlighted by Reutlinger (following David Papineau, 1993), is the fact that gases are *all* assumed to be made of a huge number of *the same kind* of particles (i.e., molecules), following the *same* mechanical laws. If what we gather under the term "gases," from a macroscopic standpoint, would be made of several distinct kinds of constituents, themselves following different laws, this explanation would crumble. For Papineau and Reutlinger, this essential appeal to a shared constitution makes this explanation an example of a "commonality strategy," which is something I agree with.

The collapse of the explanation of CP universality onto the explanation of the ideal gas law would be problematic to Batterman's goals. The latter is indeed commonly taken as one of the most classical examples of micro-reductive explanations. On the contrary, Batterman has made the former an instance of irreducible multiple realizability, thanks to the explicative value of idealizations he sees in the failure of this "common features" account. We will not engage in this issue here, but this point questions also the more modest aim I want to focus on, namely, what could be special in the way of explaining CP universality? Not only can the case of the ideal gas hardly be considered "special" in physics, but above all, it undermines the claims of statistical physicists we saw, resting upon a "deep frameshift" on the way to work with "unfaithful" models. If "one can subsume RG explanations under the commonality strategy in analogy with the statistical-mechanical explanation of the ideal gas law," (Reutlinger, 2017, p. 149) then Goldenfeld would be clearly wrong to distance it from a "traditional viewpoint" on the use of models in physics.

There is, however, one obvious difference between the common features of the ideal gas model and real gases, and the ones of the Ising model and fluids. The latter are purely macroscopic and phenomenological features, while the former are microscopic features about the constituents of the gases (i.e., molecules) and the underlying theory that is supposed to govern their behavior. This is a line of defense suggested by the examples brought together with CP universality by Batterman, using dimensional analysis, like the pendulum (see quotation of Batterman, 2002a, p. 9, already reproduced in section 2.3 above, and the detailed examination pp. 13–17; or Batterman, 2002b, pp. 29–31): "It is generally the case that for universal behavior, the microscopic details are in some sense absorbed into a finite number of phenomenological or measurable parameters." (Batterman, 2002b, p. 27) Indeed, whatever their constitution, their precise geometrical shape or their mass and its distribution in space, we can explain, thanks to the idealized model of the simple pendulum, that all real pendulums with a given length share, in the limit of small oscillations, the same period $2\pi\sqrt{l/g}$ (*l* being the length between the fixed point and the center of mass of the pendulum, and *g* the local gravitational acceleration). In the "broadly construed" universality Batterman has in mind, we could probably say that the "universality classes" of pendulums are thus defined only by their length. I will develop the case of the pendulum in section 3.2. We certainly do not start from "common features" stemming from a common microscopic theory of the constituents of the pendulum, contrary to the case of the ideal gas law. This example also seems to fit nicely with the "commonality strategy" in explaining this "universal" behavior of the pendulums, although in a different way than the ideal gas law. Lange (2009) discussed this precise example (see pp. 756–757), already arguing for a "common features" account of this explanation.

The commonality strategy is thus threatening the qualitative novelty of CP universality explanation claimed by statistical physicists from two sides, so to speak. *Something* is different in the role these commonalities play in explaining the ideal gas law and the pendulum case, but *what* exactly, and how is the explanation of CP universality actually related to these two cases?

## 3 Critical Phenomena on the Benchmark: On Different Roles Played by Analogies

In order to probe into this divergence of views while keeping in mind our main question of the claimed novelty of CP universality explanation, I propose to compare the two sides of the problem on the most systematic way possible. The statistical-mechanical explanation of the ideal gas law being an example of model-based explanation, like CP universality explanation, I prefer to consider, on the other hand, a generalization of the case of the pendulum, namely the harmonic oscillator model. It is, like the ideal gas model, one of the most classical examples of models used to produce explanations in physics. This model is used almost everywhere in macroscopic physics and allows understanding of the ubiquity of sinusoidal, isochronous, oscillations, with a period only dependent on a few generic parameters (i.e., $2\pi\sqrt{l/g}$ for the pendulum). It is unlikely that anyone would ever consider these two model-based explanations as "a new way of looking at physics." Besides, they can be taken as two ideal types of commonality strategy (as exemplified by Reutlinger, 2017, for the first one, and Lange, 2009, for the second), but with two very different ways of *using* commonalities along the explanatory process, as argued in the previous section.

I propose to use these two cases as a benchmark for our present working hypothesis of a genuine specificity of RG explanations of CP universality. If we cannot find any decisive difference between CP universality and the other two cases, it would be difficult to argue for this genuine specificity.

I will examine each of these examples in detail in the following sections; however, let us set the main ingredients shared by these two cases and the explanation of CP universality.

These three explanatory strategies subsume a diversity of real systems under one "universal" (broadly construed) law or property. In this loose sense, they *unify* different possible explanatory accounts for particular explananda, into one explanatory account of a "universal" explanandum. To discuss the relation of these three cases with the unificationist account of scientific explanation (Friedman, 1974; Kitcher, 1981) would be insightful, but is beyond the scope of this article.[7] To avoid any confusion, let us then call our three cases *universal explanations*.[8]

The search for more and more unified laws is a strong theme in physics, sometimes assumed as a goal in itself, whether or not conflated with scientific explanation (see Morrison, 2013). It is notoriously associated with micro-reduction of "upper level" phenomena to a unified physical "bottom level" (Oppenheim and Putnam, 1958). This is the approach used by the search for a "Theory of Everything" for instance (see Weinberg, 1987, for a famous defense of this trend). The idea is basically to be able to "reconstruct" the physical world starting from ultimate constituents; however, unifying physical laws can also be pursued through other ways. Pierre Duhem for instance, had a strong influence on what physicists now call "phenomenological approaches," advocating for an epistemological unification of physics, resulting from "upper-level" generalizations that avoid any ontological commitment (Duhem, 1914). Our case of the ideal gas model belongs to the first approach to unification, whereas the case of the harmonic oscillator exemplifies the second. We will examine below how CP universality is linked to these two approaches. Jordi Cat has already shown how, in condensed matter physics, the discussion on the nature of RG methods and CP universality has been deeply connected to the theme of unity in physics, though in strong opposition to the micro-reductive trend (Cat, 1998).

Another ingredient shared by these three cases is their essential use of models.[9] The growing interest in the last decades about scientific models has unfortunately not led to a standardized classification or characterization (Frigg and Hartmann, 2018); however, it can be said that the explanatory power of these models pertains generally to various analogical relationships with the physical phenomena to be explained (Hesse, 1966). Here I will take an analogy as a relation of partial similarity between relata that can be physical systems or the model itself. Thus, I prefer to analyze the role of commonalities at work in the context of model-based explanations in terms of the analogies underlying the use of the models.

Following Mary Hesse, one can distinguish between two types of analogies, and so two types of models, in physics:

> Formal models are syntactic structures; material models are semantic, in that they introduce reference to real or imaginary entities. […] Analogy relations themselves may be formal or material: they may be merely analogies of structure, such as that between a light wave and a simple pendulum, or they may introduce material similarities, as when gas particles are held to be like billiard balls in all mechanical properties relevant to Newton's laws. (Hesse, 2000, p. 299)

---

[7] Batterman (2002a) briefly compares his views about "universality" and the unificationist account, pp. 30–35.
[8] Thanks to an anonymous reviewer for this suggestion.
[9] In order to avoid any misunderstanding, even if it is difficult to give a comprehensive definition, let me say that I here mean "models" as used in practice by scientists to deal with a phenomenon, in various ways often largely independent of theories themselves. I do not mean the "models of a theory" as promoted by the so-called semantic view of theories, which is a very different meaning. See Hesse (2000) or Vorms (2018) for synoptic discussions of this distinction.

The harmonic oscillator model is clearly supported by a formal analogy, as it will be developed below. The ideal gas model, postulating imaginary point-like and non-interacting particles, belongs broadly to the second category that Hesse calls material analogy. Ernest Nagel, for his part, talks about "substantive" analogies (Nagel, 1961, pp. 110ff.). Apart from vocabulary issues, it is more appropriate to the present discussion to specify the type of analogy by talking about a *constitutive* one—that is, a relation of similarity between the *constituents* of the relata. Indeed, as we saw in section 2.4, one feature of the explanation of the ideal gas law, which is crucial for the debate, is that the essential common features of the explanation are about the microscopic constituents of the gases. We will thus consider constitutive analogy as a special type of physical analogy (called "material" by Hesse or "substantive" by Nagel).

We will see in section 3.3 how these analogies are used in the CP case. But first let me clarify the difference of approach with the recent study by Doreen Fraser (2020) about analogies in the context of RG methods. Fraser also distinguishes between formal and physical analogies at work, but questions nevertheless a different point as the present one. RG methods have developed in parallel with statistical physics and quantum field theory. Fraser is interested in the analogies built to connect these two domains and argues that *these* analogies are purely formal. Thus, they would not allow a transfer of the physical interpretation that RG methods has in statistical physics to quantum field theory. Our focus here is only on the statistical physics use of RG methods, which precisely has been given a physical interpretation, thanks to physical analogies such as those relating the Ising model to the ferromagnets.

Finally, the last main ingredient shared by these cases is the use of different idealizations to build the model and to lead to the "universal" explanandum. They are often intricated, but one can roughly distinguish, for our current purpose, abstractions (or "Aristotelian idealizations" for some authors) and, strictly speaking, idealizations. Abstractions can be defined by discarding all properties considered irrelevant and keeping in our imagination only those considered important to the process of explanation. On the contrary, some idealizations are deliberate distortions, making assumptions used in the model intentionally false (see Jones, 2005).[10] Using Batterman vocabulary, I will call asymptotic reasonings the specific idealizations built explicitly by taking the mathematical limit to zero or infinity of some quantity.

As we will see below, abstractions and formal analogies are central to the harmonic oscillator case. To fix the ideas, we will then call *Duhemian universality* the kind of universal explanation it exemplifies. Indeed, Duhem became famous for opposing, at the dawn of statistical physics, the "mechanical models" characteristic of the micro-reductive approach. He promoted instead an ideal of "abstract theories" (Duhem, 1914, 1991, especially part I, chap. IV "Abstract theories and mechanical models"), using mathematical abstraction and formal analogies as a tool of unification for physics:

> The physicist who seeks to unite and classify in an abstract theory the laws of a certain category of phenomena, lets himself be guided often by the analogy that he sees between these phenomena and those of another category. If the latter are already ordered and organized in a satisfactory theory the physicist will try to group the former in a system of the same type and form. (Duhem, 1914, p. 140; Duhem, 1991, pp. 95–96.)

---

[10] Some authors subsume "abstractions" under the label "idealizations," while others prefer to call "idealization" only the second category defined here. Whatever is the choice made for the extension of the label, I use here Jones's clarification between the two categories.

The other important point in Duhem's theses that can justify this parallel is his insistence on a unification avoiding any ontological commitment, in opposition to the mechanistic explanations based on the atomic hypothesis. The main technical tools he championed to this effect were the notion of potential and variational methods. These are tools still central to the current "phenomenological approaches" like those mobilizing the harmonic oscillator model.[11]

Before detailing the comparison between the three model-based explanatory strategies in order to try to single out those of CP, I summarize here the main answer to the question raised—namely, is there anything specific to the explanation of CP universality compared with our ideal types of micro-reductive and Duhemian universality, which could prevent Batterman's insight from "commonality strategy" objections? My proposal is that the explanation of CP universality combines constitutive analogy (like in the micro-reductive universality of the ideal gas) and formal analogy (like in the Duhemian universality of the harmonic oscillator) in an "explicative loop" that allows for an epistemic autonomy of CP universality.

3.1 The ideal gas model, an ideal type of micro-reductive universality

First, let us examine how the ideal gas model allows physicists to unify the explanation of some behavior of gases. The behavior at stake is the well-known ideal gas law we already discussed in section 2.4. All gases appear, experimentally, to behave accordingly to this relation, when they are sufficiently diluted. Statistical physicists explain the "universality" of this property by means of the construction of a model starting from a constitutive analogy, an analogy between the constituents of the gases and those of the model. Namely, the model will be *a fictional system of free and independent particles*, those fictional particles being analogues to the molecules constituting the gases. This is a perfect example of infinite idealization: we are not abstracting away from irrelevant properties, we are starting from a deliberately false assumption, i.e., that molecules do not interact, even by collision.

Different derivations of the model exist, but the development outlined below is the most commonly used (see e.g., Huang, 1987 for a classic textbook, or Kardar, 2007 for a more recent one). The model starts with the atomic hypothesis, stating that all gases are made up of the same generic kind of microscopic constituents, somehow analogous to particles, and governed by the same laws up to some parameters singling out the different gases. Since we have strong beliefs in a specific theory of these underlying constituents, we use the language of quantum mechanics to write a theoretical description of this system of free (driven only by the principle of inertia) and independent particles. The Hamiltonian operator $\hat{H}$ encodes all the physical information about the constituents, according to quantum mechanics. The Hamiltonian of non-interacting subsystems, being simply the sum of the Hamiltonians of the different subsystems, is written as the following[12]:

$$\widehat{H} = \sum_{i=1}^{N} \frac{\widehat{p_i}^2}{2m} \qquad (2)$$

with $N$ representing the number of particles, $\widehat{p_i}^2/2m$ the Hamiltonians of each particle $i$, $\widehat{p_i}$ the momentum operators for each particle $i$, and $m$ the mass of each particle.

---

[11] Even so, this "Duhemian" denomination is not intended to be historically over-interpreted, and only claims to be evocative of those particular features. Indeed, our framework of model-based explanations remains very different on other aspects of Duhem's ideal of sets of predictions purely deduced from a coherent and hierarchical theory.

Since there are different types of gases in the world that we assume to correspond to different microscopic constituents (i.e., molecules), this Hamiltonian must differ from one kind of particle to another. This can be seen here with the parameter $m$, varying from gas to gas. Another parameter $s$, the spin of the particles, is also encoded in the Hamiltonian (although invisible in this formula). This spin may also vary from gas to gas, depending on the molecules that constitute it. Then, each considered specific gas will correspond to a specific set ($m$, $s$) of parameters, while following the same formula above.

Even from this oversimplified fictional system, which is the most mathematically tractable possible, we actually cannot deduce much in practice. We need to introduce several non-deductive inferences that Batterman would call asymptotic reasoning, taking the limit of some quantities to zero or infinity. There is the thermodynamic limit, consisting of enlarging the size of the system (its volume and its number of particles) to infinity ($N \rightarrow \infty$, $N/V$ constant). This idealization has been much discussed in philosophical literature since it allows inferring thermodynamic laws from statistical-mechanical ones. The "Maxwell-Boltzmann approximation," often overlooked in philosophical discussions of the ideal gas model, is, however, probably the most important idealization for our discussion. This idealization corresponds to the limit of the large dilutions of the gas, so to a large ratio $V/N$. More precisely, it takes the limit of the large dilutions of the available quantum states $(2s+1)V/N$ of the particles, depending on the spin $s$ of the gas's particles, when compared to the "range of the quantum effects." This range is dependent on the temperature: the hotter the gas is, the bigger the range of "quantum interferences" among the particles. It is evaluated due to a quantity called the thermal de Broglie wavelength, which is inversely proportional to the square root of the mass of each particle, and to the square root of the temperature. Roughly, we can say that taking this asymptotic limit means that the dilution of the gas is large enough to ignore "quantum interferences" between its constituents. Written in a condensed form, it reads as follows (with $\hbar$ the reduced Planck constant and $k_\text{B}$ the Boltzmann constant):

$$\frac{\left(\frac{2\pi\hbar^2}{mk_BT}\right)^{3/2}}{(2s+1)\frac{V}{N}} \rightarrow 0 \qquad (3)$$

---

[12] For clarity, we here consider only the Hamiltonian of the monatomic gases. The same reasoning applies to molecular gases, with a second term in the definition of the Hamiltonian, that will not change the final result.

It is that precise idealization that allows us to eliminate the parameters *m* and *s* that singled out our different kinds of particles and gases at first. Conversely, it is the failure of this idealization in the low temperature domain that explains the appearance of the quantum gases behaviors, like superfluidity, which become dependent on the nature of the particles that constitute the gas (especially their spin, determining if the particles are bosons or fermions).[13] The unification of the behaviors of all gases at common temperatures and large dilutions thus appears to be essentially drawn from this "Maxwell-Boltzmann" limit.

Finally, we can infer the one ideal gas law after applying the thermodynamic limit. We could call this a *micro-reductive* explanatory process, especially because of a by-product of this process relating the mean kinetic energy per particle to the temperature.[14] The "universality" of the explanandum thus appears to be rooted in the asymptotic reasonings (namely, the Maxwell-Boltzmann and thermodynamic limits), which we will see again in the case of CP universality.

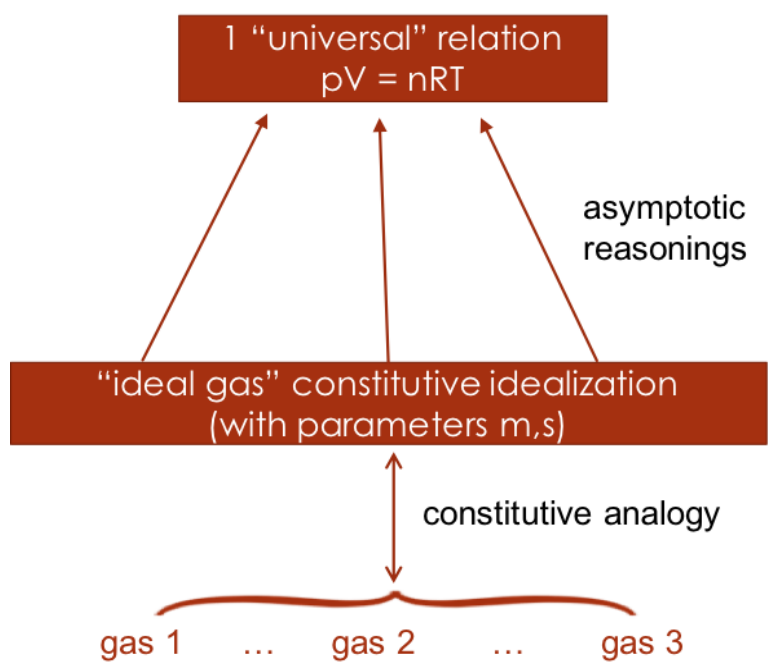

*Fig. 1* Schema of universal explanation in the ideal gas case

The general representation is illustrated in figure 1. In this diagram, similar to the following two figures, boxes represent mathematical statements, whereas elements outside boxes represent physical systems. Arrows stand for different kinds of inferences. In this case, we start from different gases. We build a constitutive idealization of these gases, namely our "ideal gas" made of free and independent particles characterized by two parameters *m* and *s* all following the same quantum mechanical equation. The explanatory use of this idealization is supported by a physical analogy relating the constituents of the different real gases (different types of molecules) and the constituents of the ideal gas (the imaginary free and independent particles, all of the same type). The arrow symbolizing this constitutive analogy is double-headed to show that the model is first constructed from assumed properties of gases (gases-to-model), but that the function of the analogy is to be able, in return, to reasonably infer other properties to the real gases, from the ones derived from the model (model-to-gases). Then, the parameters (*m*, *s*) individualizing the different constitutions of the gases in the model are eliminated by the use of asymptotic reasonings (single-headed arrow, to denote the unidirectional character of the derivation). We finally arrive at one "universal" relation independent of the parameters that initially individualized the different gases.

---

[13] For instance, the superfluidity of helium-4 below 2.17 K is linked to its spin being an integer (meaning that helium-4 is a boson). This is a transition that the other isotope, helium-3, cannot undergo because of a different spin (half-integer, meaning that helium-3 is a fermion). These two isotopes thus cannot be described by one universal law outside the domain of the Maxwell-Boltzmann idealization.

[14] I stress here that, contrary to the once influent Nagel's account of the reduction of the ideal gas law to (statistical) mechanics (see Nagel, 1961, chap. 11), this relation between temperature and mean kinetic energy is *not* an identity or even a biconditional statement. It appears as a by-product of the *non-deductive* asymptotic reasonings, valid only for specific cases such as this one, and not a prior assumption that could play the role of a "bridge principle" or a "coordinating definition" supplementing a deductive derivation of thermodynamic laws from mechanical ones.

3.2 The harmonic oscillator model, an ideal type of Duhemian universality

Let us now turn to the harmonic oscillator model, which is our example of a "universality" which is autonomous regarding the physical constitution of the phenomena considered. It generalizes the explanation of the pendulum's behavior to be applied equally to a huge number of phenomena, for instance spring-mass systems, LC electric circuits, or molecular vibrations, considered as analogs (Feynman, Leighton, and Sands, 2013, chaps. 21 and 25).

In this case, contrary to that of the micro-reductive one, we build an idealization by means of *mathematical abstraction*. Basically, based on experimental findings, we keep the phenomenological parameters regarded as useful to explain the phenomenon, and we leave aside the others, that will play no role in the explanation. Unlike the previous case, the selection of the relevant parameters is an (empirically confirmed) assumption placed on the basis of the model. For the pendulum, we can keep the angle between the string and the upright position. For the LC electric circuit, it is the charge of the capacitor. For spring-mass systems or molecular vibrations, it is the deviation length from an equilibrium position. After having written the basic equations about the behavior of these different quantities, we can identify a formal analogy among these different relations.

There is no systematic recipe to identify the right physical quantity to keep in abstraction, and to introduce in the formal analogy. There is, however, a very general mathematical result that helps explain the ubiquity of this model. Suppose we identify, in a given phenomenon, periodical variations of some parameter $x$ around a state of equilibrium $x = 0$. If this phenomenon can be described as a conservative system (i.e., that has a constant energy), through variational methods we will then be able to describe the evolution of this parameter by a motion along the abstract space of a "potential" function $V$, packing all the information about the unknown forces that might operate. Finally, we can write the finite expansion to the second order of the potential, around the equilibrium state[15]:

---

[15] Considering that the potential (and so the underlying forces) is "smooth enough" to be differentiated twice.

$$V(x) = \frac{1}{2}V''(0)x^2 + o(x^2) \tag{4}$$

with $V''$ the second derivative of the potential function, and $o(x^2)$ an arbitrary function approaching zero when $x \rightarrow 0$ and negligible compared to the previous term in the limit of the small oscillations around the equilibrium (i.e., $x \rightarrow 0$). This finite expansion corresponds in fact to the "linear domain" in terms of forces, as when one uses Hooke's law $F = kx$ for a spring. The quantity $V''(0)$, which packs all the information about the behavior of the system in the vicinity of the equilibrium, is often conventionally represented by the symbol $\omega_0^2$, as shown in figure 2.

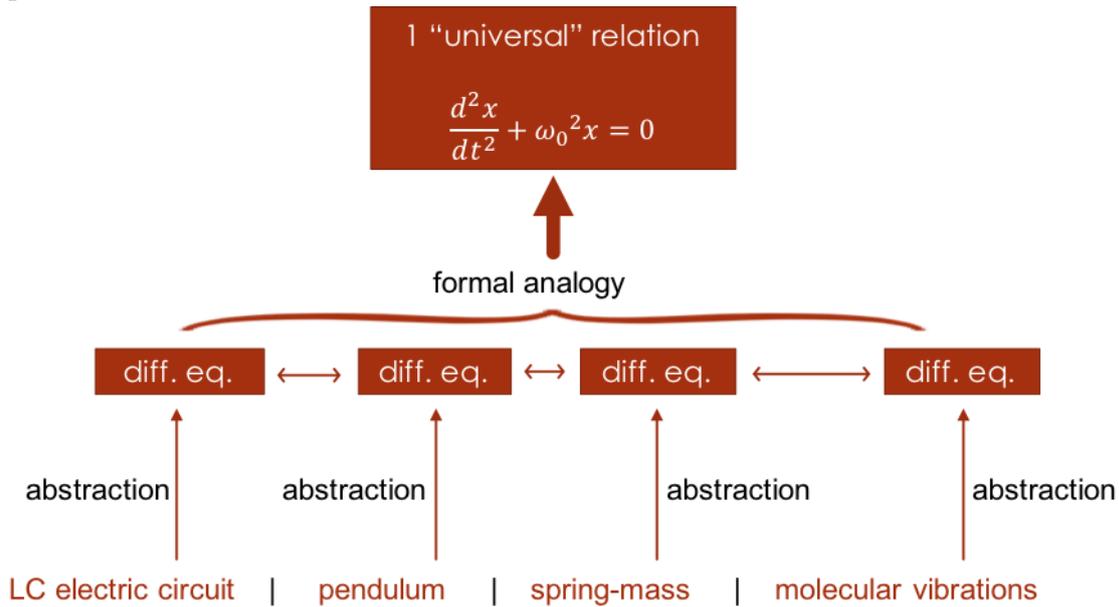

*Fig. 2* *Schema of universal explanation in the harmonic oscillator model case*

Figure 2 shows the general representation of the case of the harmonic oscillator model. We operate a mathematical abstraction on these different physical systems, possibly combined with an asymptotic reasoning, for example to make the friction tend towards zero (light single-headed arrows). We then produce differential equations which are shown to be mathematically analogous: one can systematically relate the different quantities in these equations and show that they will be identical up to these substitutions. For instance, with a spring-mass system, of force $F$ and mass $m$, and an LC circuit, of voltage $V$, inductance $L$ and intensity $I$, $V$ corresponds to $F$, $L$ corresponds to $m$, and $I$ corresponds to velocity. The formal analogy is embodied by this series of analogs, symbolized here by the double-headed arrows between the different differential equations. These separate differential equations can then be referred to by a single symbolic relation called the "harmonic oscillator equation." In this symbolic relation, the parameter $\omega_0^2$ is identified as a specific combination of the physical quantities considered in each problem, like $g/l$ in the case of the pendulum, thanks to dimensional analysis. The heavy arrow stands here for the overall abstraction-formal analogy process allowing inferring the "harmonic oscillator" relation from each of the physical systems. In particular, solving the harmonic oscillator equation then allows explaining that all these systems will display, in the limit of small oscillations, sinusoidal oscillations with a period $2\pi/\omega_0$, which is independent of the initial conditions.

Here, the universal explanation uses no constitutive analogy that could support the hypothesis of a shared constitution of all these systems. Because this analogy is conceived as "only formal," we are not inclined to infer such a shared constitution from the result of a similar behavior.[16] As Lange said for dimensional explanations, it "succeeds in unifying what separate derivations from more fundamental laws fail to unify." (Lange, 2009, p. 744)

3.3 Critical phenomena's universality: a hybrid type of universal explanation?

Finally, with CP universality, we start with a constitutive idealization of one kind of system, like with the micro-reductive strategy. In the case of uniaxial ferromagnets (e.g., nickel) it is the Ising model. It is assumed, similar to the ideal gas model, that the physical system under study consists of a collection of atomic constituents. The interesting property here, displaying a CP, is the magnetization. These constituents are then assumed to look like some "atomic magnetic moments." This is the, albeit very crude, constitutive analogy that underlies the use of the Ising model. These "atomic magnetic moments" are often seen as spins in modern interpretations of the model, but do not result from quantum mechanics.

Our fictional system will then be a lattice of $N$ different "+" and "−" magnetic moments (or "up" and "down"), the magnetic moments being supposed to acquire only two directions. These magnetic moments can interact depending on their sign. A classical mechanical description of the system is then written using a Hamiltonian function:

$$H = -J \sum_{\langle i,j \rangle} \sigma_i \sigma_j - h \sum_{i=1}^{N} \sigma_i \qquad (5)$$

with $i$ and $j$ indices standing for the $N$ sites of the lattice supposed to correspond to the atomic structure of the material, and the $\sigma_i$ and $\sigma_j$ standing for the "+1" or "−1" magnetic moments tied to the sites $i$ and $j$. The first sum is a sum on all the combinations of the nearest neighboring magnetic moments. $J$ is the "coupling constant" that quantifies the interactions between these magnetic moments. The second term of the Hamiltonian introduces the influence of an external magnetic field, $h$ being another "coupling constant" quantifying the interactions between this external magnetic field and each magnetic moment.

Like with the ideal gas model, we have thus a common description for all ferromagnets based on a hypothesis of a shared physical constitution: they are all assumed to be a collection of "microscopic magnetic moments" with two possible directions, distributed on a regular lattice, and interacting among each other and with external magnetic fields following the same law (that is, the Hamiltonian reproduced above). This common description allows individualizing the different ferromagnets by three parameters, like the parameters $m$ and $s$ of the ideal gas model: here, it is the coupling constants $J$ and $h$, and the lattice structure (determining the nearest neighbors of each magnetic moment in the first sum of the Hamiltonian).

---

[16] In this respect, what I call here a Duhemian universal explanation is very similar to what Christopher Pincock calls abstract explanations (Pincock, 2014).

With $J > 0$, if two magnetic moments point in the same direction, they will lower their energy. It is thus the favored configuration, which will compete with the random distribution of the magnetic moments at temperatures $T > 0$ K, $T$ being the control parameter. This crude model has been shown to display a phase transition at a precise critical temperature $T_c$, depending on the parameters of the model, between a magnetically ordered state (that looks like a ferromagnetic state in real materials) at $T < T_c$, and a magnetically disordered state (that looks like a paramagnetic state in real materials) at $T > T_c$. This is the main result that supports the use of this model to *represent* real ferromagnets. But this is not the way one explains the *universality* of the critical exponents in the vicinity of the phase transition.

As already roughly described in section 2.1 (see the references in this section), we need RG methods. First, we start by considering a "reduced Hamiltonian" $\mathcal{H} = -H/(k_B T)$, by dividing the coupling constants $J$ and $h$ by a factor $k_B T$, that will make the reduced Hamiltonian independent of temperature. Then, the most crucial step of the reasoning is to embed this reduced Hamiltonian into a larger *functional space*,[17] allowing any continuous distortion of the parameters of the model, including the addition of new terms. If we write $[K]$ the set of *possible* parameters of the reduced Hamiltonian, allowing extending this set of parameters, with $[K] = \{K_1 = h/(k_B T), K_2 = J/(k_B T), \ldots, K_n\}$, we can then consider a generalized Hamiltonian:

$$\mathcal{H}[K] = \sum_n K_n \Theta_n(\{\sigma_i\}) \tag{6}$$

$K_n$ being all the (reduced) possible parameters, $\{\sigma_i\}$ the degrees of freedom of the Hamiltonian (i.e., the different values of the magnetic moments of all the lattice sites $i$), and $\Theta_n$ some undetermined functions of these degrees of freedom. Our initial Hamiltonian is now only one possible $\mathcal{H}(K_1, K_2)$ among a functional space of infinite dimensions, that is, from a geometric point of view, a point on a manifold.

We then take advantage of the property of scale invariance we identified in the vicinity of the critical transition (see section 2.1), by searching for a *renormalization transformation* $\mathcal{R}_\ell$ that will *change* the Hamiltonian describing a given system to reproduce instead "what the system would look like from a scale interval $\ell$," which is the main parameter of the renormalization transformation, but *without* changing its physical properties. One can write $\mathcal{H}[K'] = \mathcal{R}_\ell(\mathcal{H}[K])$, $[K']$ being the new set of parameters that will precisely give the same global behavior after this change of scale. This transformation can be applied recursively ($\mathcal{R}_k \circ \mathcal{R}_\ell = \mathcal{R}_{k\ell}$).

Like with the harmonic oscillator case, when one needs to find the right parameter to keep in abstraction, there is no systematic recipe at this point. Each problem is a new challenge to build the appropriate renormalization transformation, and in order to succeed one needs to lean on the empirical knowledge we have of each problem. Luckily, mathematical results show that the final properties we are interested in (the universality class and the value of the critical exponents), are independent of the renormalization transformation chosen and its scale interval parameter $\ell$, as long as the procedure succeeds.

---

[17] Thanks to Annick Lesne having made this point very clear to me.

Nevertheless, we have a generic "methodological guide" to build these renormalization transformations consisting of three main steps. First, a "coarse-graining" or "decimation" of the Hamiltonian will group together several degrees of freedom. In the case of the Ising model, we group several lattice sites bearing each a "spin" to build "block spins" with a rule of majority: if the majority of the arrows symbolizing the spins are "up," then the "block-spin" will also be "up," and vice versa.[18] The Hamiltonian is then "simpler," information has been eliminated. The degrees of freedom have decreased by a factor $\ell^d$ ($d = 3$ in a three-dimensional space).

Second, since the interval between the lattice sites have changed from a length $a$ to a length $\ell a$ between "block spins," we need to apply a *rescaling* of the Hamiltonian, to "put it back to its original size." For instance, the coordinates $x$ will be substituted with $x/\ell$.

Third, we adjust (or "renormalize") the parameters $[K]$ of the Hamiltonian so that it reproduces the same physical behavior, i.e., gives the same free energy, despite the first two steps. Even if we *cannot*, in practice, determine the free energy from the Hamiltonian, we can determine how this function would change under the first two steps, and how to adjust the parameters. These new parameters $[K']$ are called *effective parameters*. Note that the *number* of effective parameters can change when applying the renormalization transformation. Even if, at first, we consider only an interaction $J$ between the nearest neighbors of the lattice, the effective behavior of the upper "block spins" does not have to comply with this rule—they can interact differently. The number of effective parameters can even go to infinity. This is precisely why we need to start by embedding our (reduced) Hamiltonian into a larger functional space and to consider any set of parameters $[K]$ whose list is not defined beforehand.

Thus, when applying RG methods, we leap from a given model, described by a given Hamiltonian, to a functional space of possible models. One does not build the renormalization transformation by working on some state realized by a specific Hamiltonian (e.g., some distribution of magnetic moments in the case of the Ising model), but on *any* possible Hamiltonian including the specific one we started with which we started.

Provided with this renormalization transformation $\mathcal{R}_\ell$, we can finally explain the universality property and show how to calculate the exact critical exponents. To that end, we need to examine the non-trivial *fixed points* of the transformation, i.e., the (reduced) Hamiltonians $\mathcal{H}^*$ such as $\mathcal{R}_\ell(\mathcal{H}^*) = \mathcal{H}^*$. One can show that these fixed points are hyperbolic, or "saddle points," namely, "attract" the Hamiltonians in their vicinity under the renormalization flow when these are along some *stable* directions, and conversely that they "repel" the Hamiltonians in their vicinity when these are along some *unstable* directions. In other words, the renormalization transformation will bring the models located in the stable directions *closer* and *drive away* the models located in the unstable directions. The universality class is thus defined as the *basin of attraction of the renormalization flow* for the stable directions of the hyperbolic fixed point. The parameters whose variations correspond to the stable directions are thus called *irrelevant* to the universality class, while the parameters whose variations correspond to the unstable directions are called *relevant*. In particular, it can be shown that the number of space dimensions *d*, and the number of components of the order parameter *n*, are usually the *only* relevant parameters.

---

[18] We follow here the block-spin Kadanoff transformation.

    How do we go back to "practical matters" from this quite abstract definition? The last part of the reasoning is to recognize that the physical properties of a system, defined from its Hamiltonian, will follow a power law in the vicinity of its critical point, with exponents directly determined by the linearization of the renormalization transformation near the fixed point. This linearization, calculated in the space of the parameters of the initial model (here, the Ising model and its reduced parameters $J/(k_B T)$ and $h/(k_B T)$), will be asymptotically similar to the eigenvalues of the renormalization transformation at the exact fixed point, which requires a functional space of a possibly infinite number of parameters $[K]$, as we saw earlier. In fact, the most important result finally is that *all* the Hamiltonians belonging to the basin of attraction of the fixed point will display asymptotically the *same* critical exponents around their critical point, since all these critical exponents will be determined by the eigenvalues of the renormalization transformation at the same fixed-point Hamiltonian $\mathcal{H}^*$. To be clear: the critical point of the Ising model is *not* the fixed point of the renormalization transformation. This fixed point is an abstract Hamiltonian independent of our initial model that requires much more complicated parameters than the crude model with which we started. The Ising model and *all the other possible models* sharing the relevant parameters (*d*, *n*) that define the universality class will, however, behave according to power laws determined only by the abstract fixed point of their basin of attraction, in the vicinity of each of their own critical point. Thus, calculating the linearization of the renormalization transformation around the critical values of the parameters of the Ising model will at the same time allow determining the critical exponents of all the possible models belonging to the universality class.

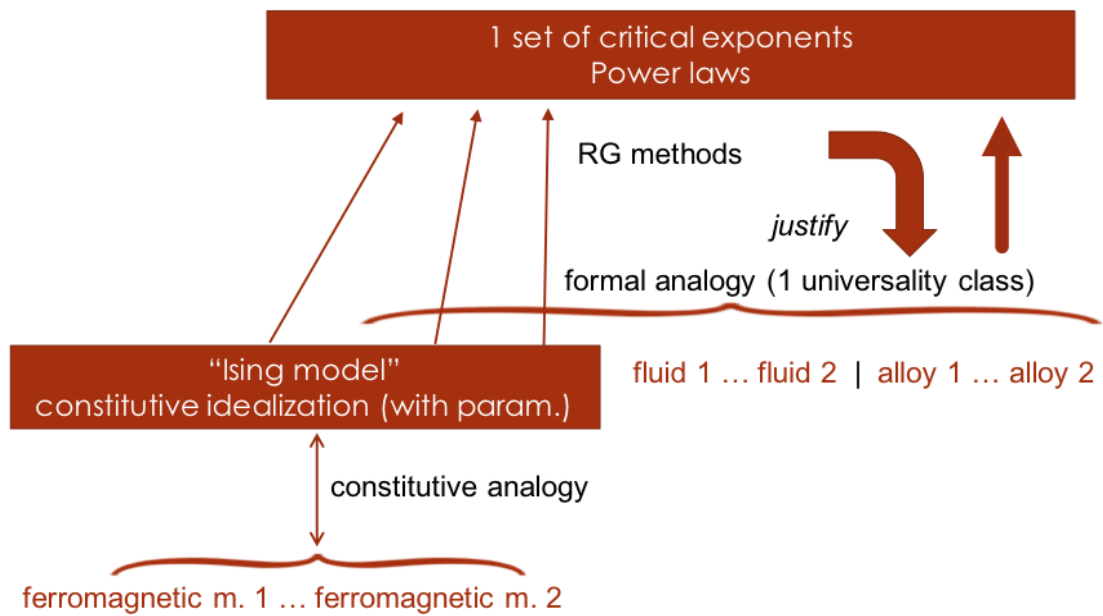

***Fig. 3*** *Schema of universal explanation in the CP case*

We can summarize the reasoning followed in the final schema illustrated in figure 3. The Ising model is an idealization that rests upon a constitutive analogy with ferromagnets (double-headed arrow), with a correspondence of some parameters like the geometric configuration of the material or the strength $J$ of the interactions between the elements. Then, the application of RG methods gives the power laws and the set of critical exponents for *this* model, eliminating the irrelevant parameters as the coupling constants or the geometric configuration of the lattice, in a very similar manner as for the case of the ideal gas, where the asymptotic reasoning gives us a single ideal gas law from a parameterized model (light single-headed arrows). In the course of doing this, however, the RG methods also give a justification for considering in the same way the critical behavior of fluids, as well as binary alloys (heavy curved arrow), because they show how all microscopic considerations about the constituents of these CP and their interactions are *irrelevant* to their critical behavior, provided they share the parameters ($d$, $n$) with the Ising model. In other words, the successful application of RG methods to the Ising model proves at the same time the independence of the exponents from any microscopic consideration and thus, their values for the whole class of possible systems. Eventually, one can know the behavior of these systems around their phase transition while making no hypothesis about their constitution. That is because RG methods show that this constitution is no longer relevant in the analogous case of the Ising model. The universality class then works like a formal analogy (curly bracket), as in the case of the harmonic oscillator: abstracting analog mathematical parameters (essentially, the order parameter) in fluids or binary alloys allows deducing (heavy straight arrow) that they share the same power laws and even the same exact value of the exponents in these laws with that of the Ising model, without any assumption about a shared constitution. Nevertheless, contrary to the case of the harmonic oscillator, the application of RG methods has given us formal grounds for this analogy, drawn from a microscopic consideration of *ferromagnetic* materials (heavy curved arrow).

The explanation of CP universality then seems to borrow from both the micro-reductive "universality" and that of the Duhemian one. Starting from an idealized constitution very specific to one kind of system, one eventually arrives at a macroscopic result independent from this constitution. It therefore ensures an *epistemic autonomy* of CP universality, both based on constitutive hypotheses but at the same time autonomous from this constitution, thanks to the "explicative loop" noticeable in figure 3.

## 4 Discussion

In a sense, Lange and Reutlinger are right to underline that the shared parameters defining the universality class are "common features" allowing the same explanation to apply to various systems. Batterman (2019) also agrees with that point. In the framework presented here, the role of these common features is played by the formal analogy. But as in the case of the harmonic oscillator, by itself, identifying these "common features" is only a "modest" explanation, in the sense that it is not grounded in our beliefs about the physical constitution of the world, represented in mathematical terms by a Hamiltonian describing some elements and the laws governing them. This would make the explanation of CP universality an explanation "from above," as in the Duhemian strategy.

Physicists historically started from such a formal analogy, before finding stronger explanations. Pierre Curie has probably been the first to note this common empirical behavior between the ferromagnets and the fluids, and to build a formal analogy between them:

> There are analogies between the function $f(I, H, T) = 0$ relative to a magnetic body and the function $f(D, p, T) = 0$ relative to a fluid. The magnetization intensity $I$ corresponds to the density $D$, the field intensity $H$ corresponds to the pressure $p$, the absolute temperature $T$ plays the same role in both cases. [...] The way in which the magnetization intensity varies as a function of temperature in the vicinity of the transformation temperature, with the field remaining constant, is reminiscent of the way in which the density of a fluid varies as a function of temperature in the vicinity of the critical temperature (with the pressure remaining constant). (Curie, 1895, p. 113, my translation)

This understanding of the universality class as a formal analogy is not limited to pre-RG methods, and continues today, as when Kadanoff says that "The analogy [...] between the liquid−gas system and the magnetic phase transition makes this magnetic susceptibility the direct analog of the compressibility." (Kadanoff, 2013a, p. 158)

This analogous behavior could well have come to be an example of "trivial" universality, to borrow Wilson's word (Wilson, 1979, p. 174). What, however, makes CP universality "non-trivial" and very different from the Duhemian strategy is that it has been explained starting from a microscopic understanding, in terms of interacting constituent particles, represented here by the use of the Ising model Hamiltonian. This fact allows us to say that CP universality *is* understood as "just resulting" from interactions between atomic constituents. To quote Kadanoff:

> the fixed-point concept describes a connection between the microscopic properties of the material, i.e. the interactions among its constituent particles and fields, and the behavior of the material on a conceptually infinite length scale. This connection is surprising and quite beautiful. (Kadanoff, 2013b, p. 35)

In a similar manner, Castiglione et al. (2008) distance RG methods explicitly from "phenomenological" approaches, which I take the harmonic oscillator to be an ideal type, calling it by contrast a "constructive" method (p. 219). In this respect, the explanation of CP universality *is also* partly a "bottom-up" explanation, as with the micro-reductive strategy.

Reutlinger and Lange, however, argue that the "explanatory power" of these "common features" defining the universality class could be distinguished "from the start" of the explanation, as when one says that "all gases are made of particles following the laws of quantum mechanics" to support the use of the ideal gas model as a start to the explanation of the universality of the ideal gas law. For instance, Lange argues that:

> The minimal model [here, the Ising model] and the target system [e.g., the fluids] share the common features that are responsible for the macrobehavior. But unlike the target system, the minimal model involves no other features that might make for added complications. It isolates the common features' contribution, showing what those common features produce in the absence of other influences and by what means they produce it. (Lange, 2015, p. 298)

It is at this point that I clearly disagree, for two main reasons. Firstly, the Ising model is indeed analogous to fluids, if one establishes a correspondence between the density on one hand and the magnetization on the other, as Curie did. But it possesses far more features than its spatial dimensionality and the number of components of its magnetization, as we saw with its Hamiltonian, also characterized by the coupling constants and the geometry of the lattice. These other features are *not* dispensable in the explanation process, even if the universality class does not depend on them at the end. The renormalization transformation is precisely built by starting with these parameters, even if we allow them to change with the transformation. In the same way, the critical exponents are calculated from the variations of these parameters under the renormalization flow. If one could completely set aside the parameters responsible for the microscopic interactions in the Ising model, it would make no sense for the physicists to say that RG methods explain universality by connecting it to microscopic interactions. It is nonetheless true that these parameters are set aside *as a result of RG methods*, at the end of the explanation process. That is why RG methods can be said to supplement the formal analogy of the universality class with a theoretical justification grounded in microscopic theories.

Secondly, we do not know from the start, for a given target system displaying a CP, which of its numerous properties will be the right order parameter to use for determining its universality class. It needs to be able to quantify the phase transition, with a non-zero value on one side of the transition. No more can be said. As Binney et al. (1992) explain, "There is no general scheme for defining order parameters; one has to consider each new physical system afresh. We can best indicate what sort of thing an order parameter is by giving a number of examples" (p. 11). This is a very important point that has been overlooked. Here, we are placed in a similar situation as with the harmonic oscillator: the choice of the right parameter to keep in order to place the phenomenon in the right universality class is a *practical matter* that needs to be confirmed. We cannot infer it from an underlying theory, and there is no rule to identify it in every instance. That means that there is no rule to determine, at first, to which universality class a given CP could belong. One can only find in textbooks lists of examples. Magnetization, in the case of ferromagnets, seems obvious, but the binary alloys, presented in section 2.1, display a far more elaborated order parameter. CP with yet more sophisticated order parameters have been identified. Thus, the order parameter is only defined, precisely, in an analogical way, without any common physical interpretation between the members of the universality class. In short, the order parameter is only defined by a *formal* analogy. Using physicists' terminology, one can say that the order parameter is identified on a phenomenological basis.

I argue that the deep irrelevance of the microscopic description of a CP to determine its critical behavior is linked to this point. If the different order parameters could all be considered as *physically* similar in terms of microscopic interpretation, the "commonality strategy" could indeed subsume the explanation of CP universality, as it does with the ideal gas and the harmonic oscillator. The strong specificity of CP, however, is, I argue, the consequence of this explicative loop produced by the interplay between a constitutive analogy giving a microphysical interpretation of the universality on one hand and a purely formal analogy that allows us, on the other hand, to extend the results gained on a specific model, to phenomena whose microscopic description can be completely unrelated. In brief, we draw from microscopic physics grounds to rule out the very microscopic physics of CP–something one could call *epistemic autonomy*.

Let us finally go back to Batterman's view about the indispensability of RG methods for explanation. If we just focus on using the empirically confirmed delimitation of each universality class in order to explain the behavior of some specified system, we only need the formal analogy, and so the explanation complies with the commonality strategy. Likewise, if we just focus on using the Ising model in order to explain the behavior of some ferromagnet, from a microscopic viewpoint grounded in well-confirmed theories, we only need the constitutive analogy, and so the explanation also complies with the commonality strategy. But if we want to explain, from a microscopic viewpoint grounded in well-confirmed theories, why *all* CP displaying a scalar order parameter will exhibit the same critical exponents as the Ising model whereas presumably sharing *no* microscopic commonalities with it, we *need* to perform the asymptotic reasoning used in the form of RG methods. This is the indispensable element allowing explaining, starting from a (microscopic) commonality, another (phenomenological) commonality that cannot be reduced to the former. In this sense, I follow Batterman's view; although, I argue that it is only by disentangling two different uses of commonalities at work (in the form of the Duhemian strategy or the micro-reductive one) that one can identify at which point RG methods are precisely indispensable, and what makes this explanation singular.

# 5 Conclusion

We followed Batterman's insight that some deliberate distortions of models are responsible for the capability of the property of "universality" to unify a large class of phenomena, regardless of their distorted microscopic description. Restricting Batterman's account of "universality" in its core example, i.e., critical phenomena, this article shows that assessing the differences between CP universality and two paradigmatic cases of "commonality strategy" of explanation allows avoiding the objections raised by proponents of this approach, and to single out the claimed specificity of this explanation.

Taking the ideal gas model and the harmonic oscillator model as two benchmarks for CP reveals the importance of the different roles played by the analogies underlying the use of the models. In the case of the ideal gas model, there is a constitutive analogy about the physical constituents of the gases that supports an idealization. In the case of the harmonic oscillator model, the explicative unification is essentially brought by a formal analogy, supported by the means of mathematical abstractions. The case of CP universality is then shown to be a combination of these two explanatory strategies. The RG methods need to start from a constitutive analogy with some physical system, but eventually justify a formal analogy between the initial idealized model, and any systems sharing nothing but a few abstracted structural properties. This "explicative loop" ensures a specific epistemic autonomy for CP universality.

# Acknowledgments

I gratefully thank Chris Pincock, Annick Lesne, Vincent Ardourel, Henri Galinon and Sébastien Gandon for their helpful feedback on the manuscript. I also thank the two anonymous reviewers for their constructive remarks. A previous version of the paper has been presented at the "Emergence: Conceptual and Philosophical Aspects" workshop of the Dutch Institute for Emergent Phenomena, Amsterdam, May 2019, and benefited from discussions with the audience. I finally acknowledge the assistance of Taya Flaherty for language editing. This work has been funded by the Regional Council of Auvergne and Université Clermont Auvergne (France).